# Photoacoustic Identification of Laser-induced Microbubbles as Light Scattering Centers for Optical Limiting in Liquid Suspension of Graphene Nanosheets

Qiuhui Zhang[a,c,†] Yi Qiu[b,c,†] Feng Lin[c,d] Chao Niu[e] Xufeng Zhou[f] Zhaoping Liu[f] Md Kamrul Alam[g] Shenyu Dai[c,h] Jonathan Hu[e] Zhiming Wang[d] Jiming Bao[*,d,g]

[a]Department of Electrical Information Engineering, Henan University of Engineering, Xinzheng, Henan 451191, China

[b]School of Sciences, Southwest Petroleum University, Chengdu 610500, China

[c]Department of Electrical and Computer Engineering, University of Houston, Houston, Texas 77204, USA

[d]Institute of Fundamental and Frontier Science, University of Electronic Science and Technology of China, Chengdu, Sichuan 610054, China

[e] Department of Electrical & Computer Engineering, Baylor University, Waco, Texas 76798, USA

[f]Key Laboratory of Graphene Technologies and Applications of Zhejiang Province Ningbo Institute of Materials Technology and Engineering, Chinese Academy of Sciences Ningbo, Zhejiang 315201, China

[g]Materials Science and Engineering, University of Houston, Houston, Texas 77204, USA

[h]College of Electronics & Information Engineering, Sichuan University, Chengdu, Sichuan 610064, China

**Corresponding Authors**

*E-mail: jbao@uh.edu

**Abstract**

Liquid suspensions of carbon nanotubes, graphene and transition metal dichalcogenides have exhibited excellent performance in optical limiting. However, the underlying mechanism has remained elusive and is generally ascribed to their superior nonlinear optical properties such as nonlinear absorption or nonlinear scattering. Using graphene as an example, we show that photo-thermal microbubbles are responsible for the optical limiting as strong light scattering centers: graphene sheets absorb incident light and become heated up above the boiling point of water, resulting in vapor and microbubble generation. This conclusion is based on direct observation of bubbles above the laser beam as well as a strong correlation between laser-induced ultrasound and optical limiting. *In-situ* Raman scattering of graphene further confirms that the temperature of graphene under laser pulses rises above the boiling point of water but still remains too low to vaporize graphene and create graphene plasma bubbles. Photo-thermal bubble scattering is not a nonlinear optical process and requires very low laser intensity. This understanding helps us to design more efficient optical limiting materials and understand the intrinsic nonlinear optical properties of nanomaterials.

**KEYWORDS:** bubble scattering, photoacoustics, Raman, graphene, optical limiting

**Introduction**

Optical limiting (OL) of a medium describes drastically reduced transmission for high intensity laser beams. Because this property can be used to prevent potential laser damage, OL has attracted a lot of attention since the invention of the laser.[1] However, it has been challenging in identifying or developing suitable materials to exhibit OL at a controlled threshold of laser intensity.[1-5] Due to the same reason, the mechanism of OL has also remained an active research field. Optical limiting is generally regarded as a nonlinear optical phenomena because it can be induced by nonlinear absorption or scattering.[1-5] Recently, atomically thin 2D nanomaterials have emerged as novel optoelectronic materials with strong nonlinear optical properties.[6-10] As an important application, OL has been reported in liquid suspensions of graphene,[3, 5, 11-19] graphene oxide (GO),[3, 18, 20-29] transition metal dichalcogenides (TMD),[4, 30-36] and black phosphorous.[2, 37-40] Some OL were observed even with CW lasers.[28, 29, 41, 42]

However, as in previous OL materials, the mechanism of OL in these 2D nanomaterials has not been clearly identified, and more importantly, many proposed mechanisms are either confusing or misleading as to whether the OL is originated from the intrinsic nonlinear optical properties of nanomaterials.[11-13, 16, 17, 20-22, 30-32, 34, 37, 43-50] Based on the observed strong scattered light, the majority of reported works attribute OL to nonlinear scattering.[1, 3, 11-13, 16, 32, 37, 43, 45-49] Bubbles were suggested as possible nonlinear scattering centers, but there is no direct experimental proof, and it was still not clear whether the bubbles were generated by plasma breakdown of nanomaterials[13, 48, 49] or evaporation of solvents,[3, 12, 13, 16, 30-32, 37, 43, 45, 47, 48] A serious confusion arises when many people still associate OL with nonlinear optical properties of nanomaterials by simply referring it as nonlinear scattering despite the proposed bubble scattering mechanism.[11-13, 16, 32, 37, 43, 45, 47] To make the situation worse, nonlinear optical coefficient $\chi^3$ of nanostructures was also calculated from Z-scan measurement before identifying the OL mechanism,[3, 11-13, 16, 17, 21, 22, 28, 32, 37, 44, 47-49, 51, 52] because if the scattering comes from bubbles, it has nothing to do with $\chi^3$.

In this work, we chose graphene as a representative nanomaterial to investigate the mechanism of OL.[2-5, 11-14, 16, 17, 22, 30] Using direct imaging of bubbles and the correlation between photoacoustic signals and optical limiting, we conclude that laser-induced microbubbles are responsible for the sudden drop in optical transmission. We also point out that this is the most effective method to achieve optical limiting at low laser intensity, and bubble-induced optical limiting cannot be regarded as nonlinear scattering because it is not directly related to the nonlinear optical property of nanomaterials. This conclusion is applicable to other low dimensional materials. The understanding and clarification of OL mechanism helps to design more efficient optical limiter and explore the intrinsic nonlinear optical properties of nanomaterials.

**Results and discussion**

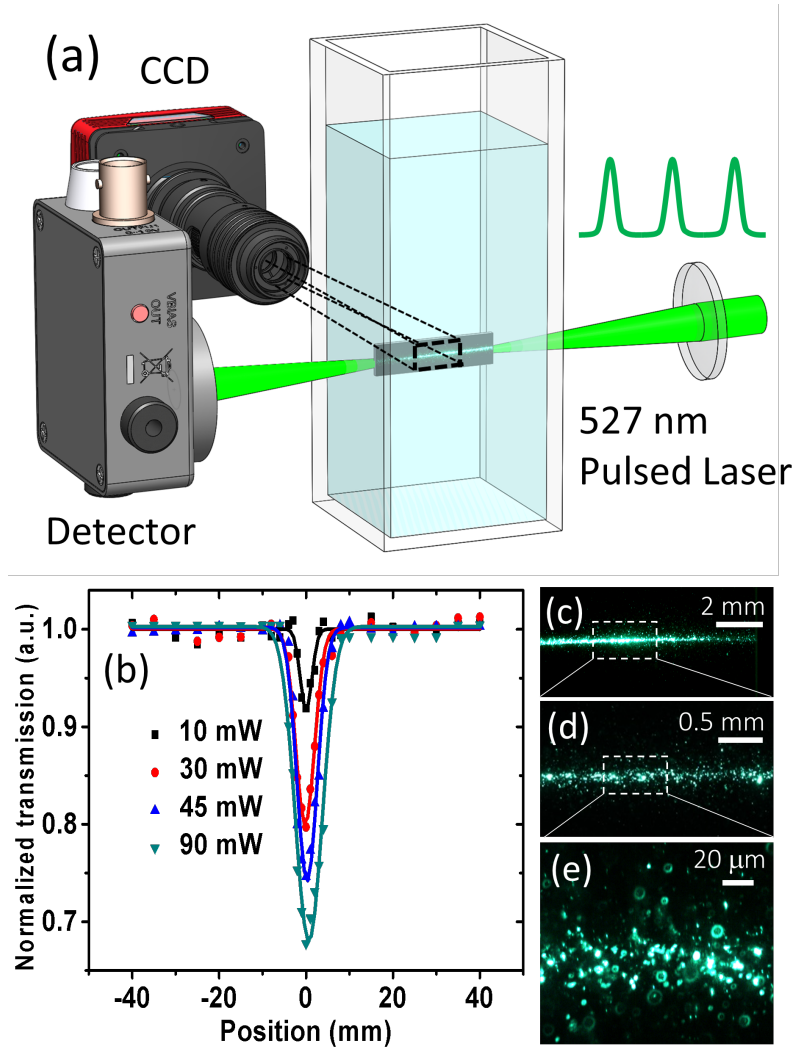

Fig. 1. Open-aperture optical limiting Z-scan experiment with direct imaging of light scattering centers. (a) Schematic for the experimental setup. (b) Z-scan curves of graphene suspension with different laser powers. (c-e) Optical images of scattering centers with increasing magnifications.

Graphene nanosheets were synthesized via electrochemical exfoliation of highly oriented pyrolytic graphite (HOPG) in $K_2SO_4$ salt solution.[53] After filtration and ultra-sonication in NMP for 2 hours, the average size of the graphene was 1.5 μm with a thickness of 2.4 nm.[54, 55] Fig. 1a shows our initial experimental setup to investigate the mechanism of OL. In addition to the traditional open-aperture Z-scan configuration, we used a high-speed video camera coupled with a microscope objective lens. This setup can not only monitor the scattered light like a photodetector used in previous z-scan experiments,[11, 12, 16, 32, 37, 45] but also directly image the motion of graphene sheets and emerging bubbles. Here, a 527-nm pulsed laser (150 ns pulse width, 1 kHz repetition rate) was focused with a 10 cm focal length lens on a cuvette, which was filled with graphene suspension in deionized water (DIW). Fig. 1b shows normalized Z-scan transmission at increasing laser power. Optical limiting can be clearly confirmed by the sudden drop in transmission near the focus point of the laser beam. As usual, a stronger optical limiting, i.e., lower optical transmission, is achieved with an increasing laser power.

The stronger scattering from the focused laser spot is obvious from Fig. 1c. To find out whether the strong scattering originated from bubbles, we zoomed in the camera to obtain higher resolution pictures (Fig. 1d-e). Bubble-like ring objects could be seen in the Fig. 1d. However, we quickly realized that these were not bubbles, they were the same bright spots as in low resolution, appearing as bubbles when bright spots were out of focus. Because the lateral sizes of graphene sheets are typically in the range of 500 nm to a few micrometers, larger than the wavelength of visible light, graphene sheets can scatter light and appear as bright spots. In other words, both graphene and bubbles can scatter light, it is difficult to distinguish them with this direct optical imaging.

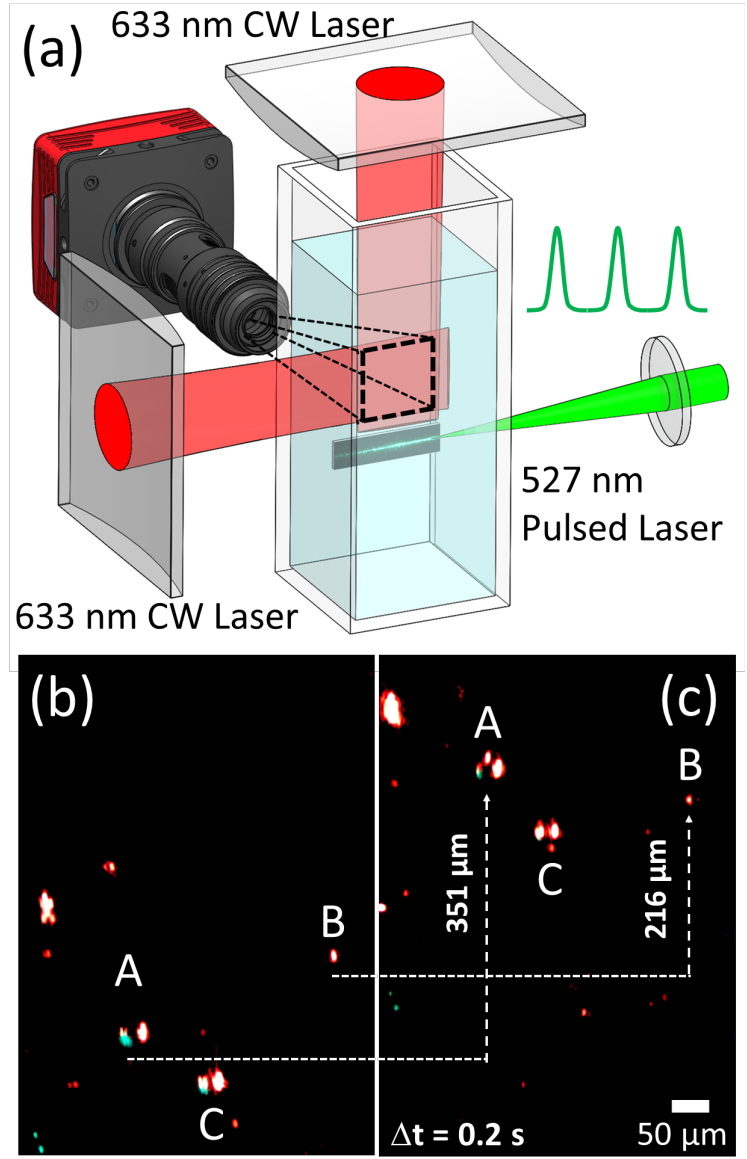

Fig. 2. Experimental setup to observe and identify photothermal bubbles above the laser beam. (a) Experimental setup. A light sheet is created above the laser beam with a 633-nm laser. (b-c) Identification of bubbles from their shapes, large sizes and faster speeds moving to the water surface.

To distinguish graphene scattering from potential bubble scattering and to obtain a clear image of bubbles, we designed a new configuration. As shown in Fig. 2a, we focused two perpendicular 633-nm laser beams with two cylindrical lenses to create a light sheet above the 527-nm laser beam, and monitored the area for a possible bubble in this thin region.[56, 57] There are several reasons for this detection configuration. First, a thin light sheet allowed us to detect bubbles only in the light sheet and made bubbles outside the light sheet invisible so that rings-like bubble artifacts can be avoided. Second, the sheet provided a large space for us to study the dynamics of possible bubbles. Third, we expected to observe larger size bubbles as microbubbles merged when drifting upward due to the buoyant force. The rationale behind this configuration is that if there were no micro-bubbles generated in the laser beam, we should not observe any bubbles above the beam.

As expected, fewer scattering centers were observed due to the thin light sheet. Fig. 2b-c show two consecutive snapshots separated by 200 ms. No ring-shaped scattering centers were observed, confirming that they were caused by out-of-focus of the camera. Now there were two types of scattering centers: large-sized centers with two or three bright points together such as A and C versus an individual, dimmer point such as B. A and C were bubbles with diameter of 30 μm, their two left and right bright spots were due to scattering of 633-nm laser in the horizontal directions, while a relatively weak point was due to the scattering of light in the vertical direction. The scattering by a single graphene sheet produced a single isolated spot, such as point B. This identification of bubbles from graphene sheets can be further confirmed by their higher upward moving speed: The bubble A moved up by 351 μm while the graphene B only traveled 261 μm in 0.2 second. This is because the motion of graphene sheet was driven by the fluid convection, which was induced by local laser heating. Bubbles moved faster because of the additional buoyant force. Note that bubbles were only observed above the focused laser beam when optical limiting occurred. These observations indicate that microbubbles must be produced in the laser beam during the optical limiting.

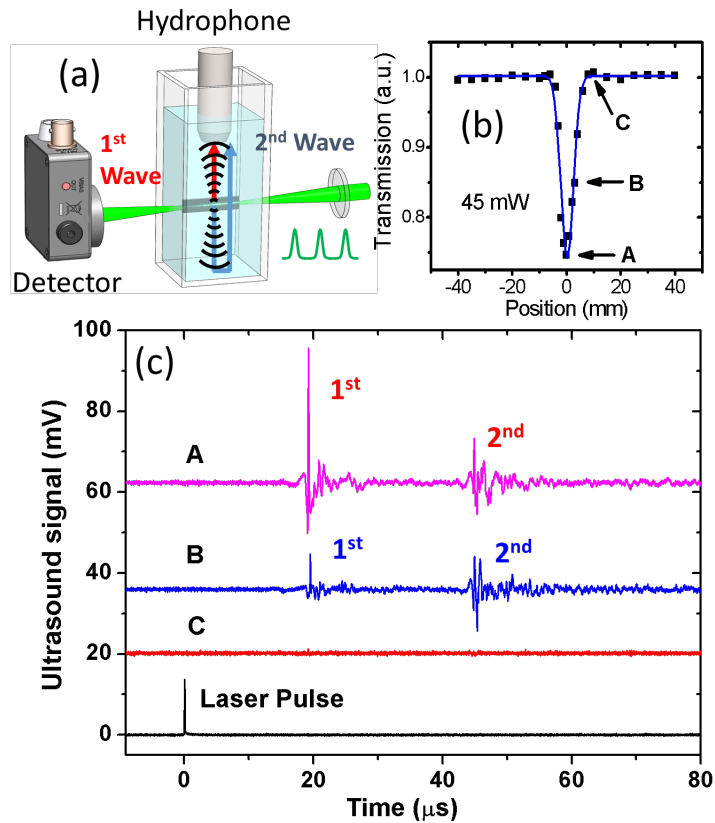

Fig. 3. Experimental setup to correlate optical limiting with photoacoustic signal of bubbles. (a) Schematic of experimental setup. (b) The same Z-scan curve of graphene suspension as in Fig. 1b under 45 mW laser. (c) Ultrasound signals at different positions of optical limiting Z-scan curve. The second ultrasound pulses around 45 μs were due to reflection from the bottom of the cuvette. The laser pulse is also included to mark the beginning of laser excitation.

Having observed bubbles during OL, we employed a new technique to further confirm the generation of microbubbles and identify them as light scattering centers. Fig. 3a shows the new experimental design where a hydrophone was used to detect ultrasound.[56, 57] The purpose was to hear laser-induced micro-bubbles instead of seeing them. To establish a tight correlation between ultrasound generation and optical limiting, we performed OL first, obtained the Z-scan curve in Fig. 3b, and then chose three positions A, B, and C. The position A exhibited the strongest OL, while the position C had no OL. Fig. 3c shows the corresponding ultrasound traces. It can be seen that the strongest ultrasound was observed at position A, and the ultrasound was too weak to be detected at position C. We want to point out that under nanosecond laser excitation, graphene can produce ultrasound through thermal expansion at position C, but that ultrasound signal will be dramatically enhanced with the microbubbles that were generated at position A.[58-60]

A pre-condition for bubble generation is that the temperature of graphene sheets must become higher than the boiling point of water under laser excitation. To estimate the rise of temperature due to laser absorption during OL, we used the same excitation laser to measure the Raman shift of graphene. Because of relatively long interaction time (~150 ns), this Raman will reflect an average temperature of graphene during the laser irradiation. Fig. 4a shows the Raman-OL experimental setup, and Fig. 4b shows Raman spectra of graphene sheets under the same laser powers of 10 and 45 mW as in Fig. 1. A Raman shift of nearly 3 cm$^{-1}$ was observed, corresponding to a rise of 180 °C,[61] which brings the sheets above the boiling point of water. This proves that the temperature of graphene became high enough to generate vapor on its surface. However, this temperature rise was certainly not high enough to vaporize graphene and create a micro-plasma which could also become a light scattering center.[62-64] Such mechanism can be further ruled out since no blackbody radiation in the visible wavelength was observed.

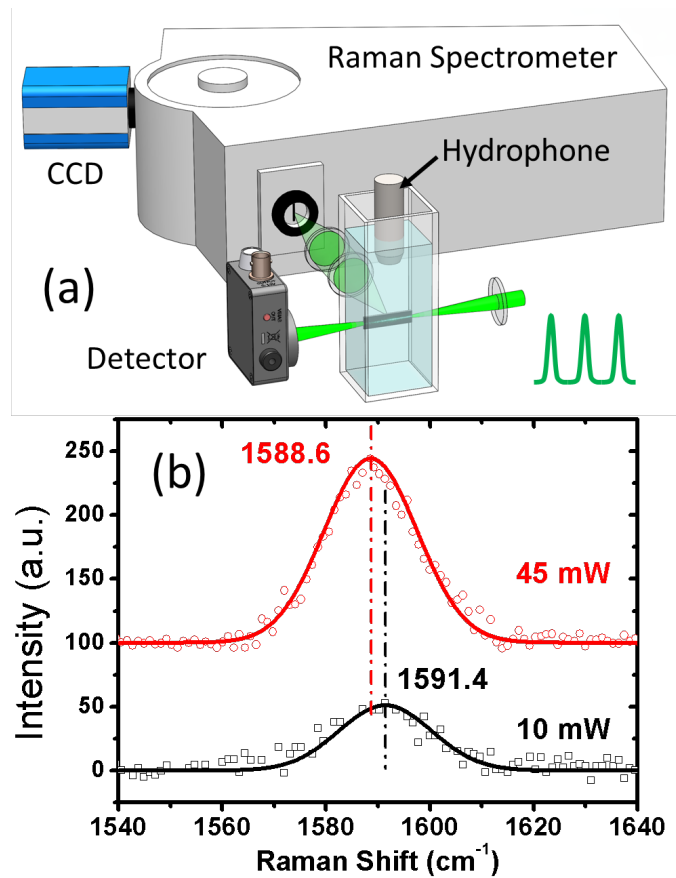

Fig. 4. Measurement of graphene temperature using Raman scattering. (a) Schematic for the experimental setup. (b) The Raman spectra of graphene under 10 mW and 45 mW of laser powers.

Based on the above observations and discussions, we can now depict each step of OL. As shown in Fig. 5a, graphene sheets absorbed incident laser energy and became hot, vaporizing surrounding water and producing micro-bubbles,[65, 66] which in turn strongly scatter incident laser and reduce its transmission. Microbubbles were proposed as light scattering centers in optical limiting,[11-13, 32, 37, 43, 45, 47] as a bubble can create total reflection of light, but the scattering properties of bubbles were not quantitatively investigated. Here we use FDTD to obtain the transmission of light through a bubble, assuming that a 0.8- μm-diameter graphene sheet is located in the center of the bubble with different sizes, as shown in Fig. 5a. Figs. 5 b-e reveal that bubbles can greatly scatter light, as 10 cascaded bubbles with 5-μm diameter can reduce the transmission to 60 %.

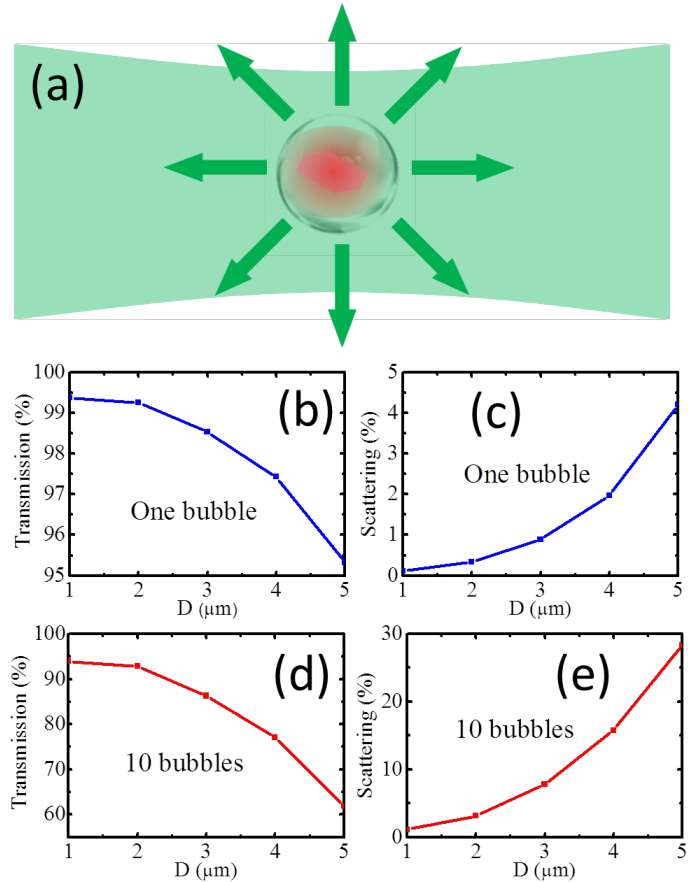

Fig. 5. Scattering of light by laser-induced bubbles. (a) Cartoon depicting laser heating of graphene flake creates a bubble scattering center. (b, d) Optical transmission induced by one (ten) bubble vs the diameter of a bubble. (c, e) The scattering induced by one (ten) bubble vs the diameter of bubble.

Laser induced ultrasound was used to investigate OL in carbon black and $TiS_2$, and a similar correlation was obtained.[31, 48, 67] However, thermal expansion of nanomaterials or solvents, instead of bubbles, were considered to generate the photoacoustic signal, and OL was attributed to nonlinear absorption of carbon black or $TiS_2$.[31, 48, 67] Some of these Z-scan experiments were performed using close aperture,[28, 68] in which the thermal lens effect of the solvents could also play an important role.[69, 70] Again, this is not a nonlinear optical property of nanomaterials. Nonlinear scattering of nanomaterials was traditionally referred to their intrinsic nonlinear optical properties. Under a higher laser intensity, the optical refractive index of suspended nanomaterials increases, resulting in larger index mismatch with their surrounding media and stronger light scattering.[71-74] Phenomenologically, the reduced transmission in Z-scan appears as a nonlinear effect, but it is misleading to simply call it nonlinear scattering,[11-13, 16, 32, 37, 43, 45, 47] and the calculation of $\chi^3$ based on Z-scan alone without knowing the underlying mechanism should be avoided.[3, 11-13, 16, 17, 21, 22, 28, 32, 37, 44, 47-49, 51, 52]

**Conclusions**

In conclusion, we have designed and developed a series of experiments to prove that laser induced bubbles are responsible for the observed optical limiting in graphene suspension. The same techniques and mechanism are applicable to other 2D nanomaterials and even carbon nanotubes.[47-50, 52] Bubble scattering is not a nonlinear optical process, so the mechanism of such OL cannot be simply called nonlinear scattering. An accurate understanding and identification of the mechanism of optical limiting is crucial for the design of effective laser protection media and exploration of optical application of nanomaterials' intrinsic properties. Because of the low laser intensity required to generate microbubbles, it is possible to use graphene to design broadband efficient optical limiting devices.

**Author Contributions**

†These authors contributed equally to this work.

**Notes**

The authors declare no competing financial interest.

**ACKNOWLEDGMENTS**

Q.H. Zhang acknowledges support from National Natural Science Foundation of China (No.61805071, 61605186). J.H. acknowledges support from U.S. National Science Foundation (ECCS-1809622). J.M. Bao acknowledges support from Welch Foundation (E-1728) and National Science Foundation (EEC-1530753).